\documentclass[aps,prd,twocolumn,preprintnumbers,amsmath,amssymb,
superscriptaddress,
,floatfix]{revtex4}

\usepackage{graphicx}
\usepackage{dcolumn}
\usepackage{bm}
\usepackage{epsfig}
\usepackage{graphics}
\usepackage{latexsym}

\def\beq{\begin{equation}}
\def\eeq{\end{equation}}
\def\bea{\begin{eqnarray}}
\def\eea{\end{eqnarray}}
\newcommand{\beqs}{\begin{subequations}}
\newcommand{\eeqs}{\end{subequations}}

\newcommand{\cref}[1]{Ref.~\cite{#1}}

\newcommand{\vev}[1]{\left<#1\right>}

\newcommand{\hh}{{\ensuremath{I{\kern-2.6pt h}}}}
\newcommand{\bhh}{{\ensuremath{\bar{I{\kern-2.6pt h}}}}}

\begin{document}

\preprint{UT-STPD-17/01}

\title{Light sterile neutrinos, dark matter, and new 
resonances \\ in a $U(1)$ extension of the MSSM}

\author{A. Hebbar}
\email{aditmh@gmail.com}
\affiliation{Bartol Research Institute, Department of Physics and 
Astronomy, University of Delaware, Newark, DE 19716, USA}
\author{G. Lazarides}
\email{lazaride@eng.auth.gr} \affiliation{School of Electrical and
Computer Engineering, Faculty of Engineering, Aristotle University
of Thessaloniki, Thessaloniki 54124, Greece}
\author{Q. Shafi}
\email{shafi@bartol.udel.edu}
\affiliation{Bartol Research Institute, Department of Physics and 
Astronomy, University of Delaware, Newark, DE 19716, USA}

\date{\today}

\begin{abstract}
We present $\psi'$MSSM, a model based on a $U(1)_{\psi'}$ extension 
of the minimal supersymmetric standard model. The gauge symmetry 
$U(1)_{\psi'}$, also known as $U(1)_N$, is a linear combination of 
the $U(1)_\chi$ and $U(1)_\psi$ subgroups of $E_6$. The model 
predicts the existence of three sterile 
neutrinos with masses $\lesssim 0.1~{\rm eV}$, if the 
$U(1)_{\psi'}$ breaking scale is of order 10 TeV. Their 
contribution to the effective number of neutrinos at 
nucleosynthesis is $\Delta N_{\nu}\simeq 0.29$. The model can 
provide a variety of possible cold dark matter candidates 
including the lightest sterile sneutrino. If the $U(1)_{\psi'}$ 
breaking scale is increased to $10^3~{\rm TeV}$, the 
sterile neutrinos, which are stable on account of a $Z_2$ symmetry, 
become viable warm dark matter candidates. The observed value of 
the standard model Higgs boson mass can be obtained with relatively 
light stop quarks thanks to the D-term contribution from 
$U(1)_{\psi'}$. The model predicts diquark and diphoton resonances 
which may be found at an updated LHC. The well-known $\mu$ problem 
is resolved and the observed baryon asymmetry of the universe can 
be generated via leptogenesis. The breaking of $U(1)_{\psi'}$ 
produces superconducting strings that may be present in our galaxy. 
A $U(1)$ R symmetry plays a key role in keeping the proton stable 
and providing the light sterile neutrinos.
\end{abstract}

\maketitle

\section{Introduction}

$E_6$ grand unified theory (GUT) \cite{E6} contains 
two especially interesting 
maximal subgroups for model building, namely $SU(3)^3$ and 
$SO(10)\times U(1)_{\psi}$. Supersymmetric (SUSY) models based on 
$SU(3)^3$, sometimes referred to as trinification models, have 
been extensively discussed in the literature. For instance, in 
SUSY $SU(3)^3$, mecha\-nisms have been proposed to resolve 
\cite{mutri} the minimal supersymmetric standard model (MSSM) 
$\mu$ problem or make \cite{protontri} the proton essentially 
stable.

The subgroup $SO(10)\times U(1)_{\psi}$ of $E_6$ can be decomposed 
further, via $SU(5)$, to the MSSM gauge symmetry group accompanied 
by $U(1)_{\chi}\times U(1)_{\psi}$ \cite{chipsi,E6SSM}. One 
intriguing combination of these two $U(1)$'s, denoted here as 
$U(1)_{\psi'}$ (also known as $U(1)_N$ \cite{chipsi} in the 
literature), is assumed \cite{Athron} here to be broken at a scale 
at least an order of magnitude greater than the TeV scale of soft 
SUSY breaking. We refer to this extension of the MSSM accompanied by 
$U(1)_{\psi'}$ as $\psi'{\rm MSSM}$. The well-known right handed 
neutrino contained in the matter 16-plet of $SO(10)$ transforms as 
a singlet under $U(1)_{\psi'}$. This enables the three right handed 
neutrinos to acquire large masses, so that the standard seesaw 
scenarios can apply and high scale leptogenesis \cite{lepto} 
can be realized \cite{sarkar}. Note that the subscript $\psi'$ 
reiterates the essential role played by $U(1)_{\psi'}$ in 
resolving the MSSM $\mu$ problem.

Our $\psi'{\rm MSSM}$ model employs in an essential way a $U(1)$ R 
symmetry such that dimension five and higher dimensional operators 
potentially causing proton decay are eliminated. The MSSM $\mu$ 
problem is also resolved and the usual lightest SUSY particle 
of MSSM remains \cite{dm} a 
compelling dark matter candidate. More intriguingly perhaps, the 
model predicts that the three $SO(10)$ singlet sterile neutrino 
matter fields that it contains can only acquire tiny masses, 
on the order of $0.1~{\rm eV}$ or less if $U(1)_{\psi'}$ is broken 
around 10 TeV.  We estimate that for this case the effective number 
of neutrinos during nucleosynthesis is changed by $\simeq 0.29$. 
The lightest sterile sneutrino as well as two more particles, 
which are stable on account of discrete symmetries, can, under 
certain circumstances, be additional cold dark matter candidates.  

If the breaking scale of $U(1)_{\psi'}$ is increased to 
$10^3~{\rm TeV}$ or so, the sterile neutrinos, 
which happen to be stable on account of a $Z_2$ 
symmetry, become plausible candidates for keV scale warm 
dark matter.

The contribution of the D-term for $U(1)_{\psi'}$ to the mass 
of the lightest CP-even neutral Higgs boson of the MSSM can be 
appreciable leading, in the so-called decoupling limit, to the 
observed value of $125~{\rm GeV}$ with relatively light stop 
quarks.  

In addition to the $Z'$ gauge boson associated with the breaking 
of the $U(1)_{\psi'}$ gauge symmetry, the model predicts the 
existence of diphoton 
\cite{theory} and diquark \cite{diquark} resonances with masses in 
the TeV range. A high luminosity or high energy ($33~{\rm TeV}$) LHC 
upgrade may be able to find them. Note that the $U(1)_{\psi'}$ 
breaking produces superconducting strings \cite{witten} which 
presumably survived inflation and should be present in our galaxy. 
If the breaking scale is not too high, a 100 TeV collider may be 
able to make these strings, which definitely would be exciting.

The layout of our paper is as follows. In Sec.~\ref{sec:model}, we
introduce the model with its field content, symmetries, and couplings.
In Sec.~\ref{sec:details}, we analyze the details of the spontaneous 
symmetry breaking of $U(1)_{\psi'}$, while in Sec.~\ref{sec:EWSB} we
discuss the spontaneous breaking of the electroweak symmetry. 
Sec.~\ref{sec:diphoton} is devoted to the diphoton excess and 
Sec.~\ref{sec:analysis} to the presentation of a numerical example.
In Sec.~\ref{sec:sterile}, we study the sterile neutrinos. The 
possible composition of dark matter in the universe is presented in 
Sec.~\ref{sec:dm} and our conclusions are summarized in 
Sec.~\ref{sec:sum}. 

\section{The model} 
\label{sec:model}
We consider a SUSY model based on the gauge group $G_{\rm SM}\times
U(1)_{\psi'}$, where $G_{\rm SM}=SU(3)_{\rm c}\times SU(2)_{\rm L}
\times U(1)_Y$ is the standard model (SM) gauge group. The 
GUT-normalized generator $Q_{\psi'}$ of the extra local 
$U(1)_{\psi'}$ symmetry is given by
\begin{equation}
Q_{\psi'}=\frac{1}{4}(Q_{\chi}+\sqrt{15}Q_{\psi}),
\end{equation}  
where $Q_{\chi}$ is the GUT-normalized generator of the 
$U(1)_{\chi}$ subgroup of $SO(10)$ which commutes with its $SU(5)$ 
subgroup and $Q_{\psi}$ is the GUT-normalized generator of the 
$U(1)_{\psi}$ subgroup of $E_6$ which commutes with its $SO(10)$ 
subgroup. The $U(1)_{\psi'}$ symmetry is to be spontaneously 
broken at some scale $M$ and we prefer to implement this breaking 
by a SUSY generalization of the well-known Brout-Englert-Higgs 
mechanism. 

The important part of the  superpotential is 
\bea
W &=&y_uH_{u}^{1}qu^c + y_dH_{d}^{1}qd^c + y_{\nu}H_{u}^{1}l\nu^c +
y_{e}H_{d}^{1}le^c
\nonumber
\\
& & +\frac{1}{2}M_{\nu^c}\nu^c\nu^c+\lambda_{\mu}^{i}
NH_{u}^{i}H_{d}^{i}+\kappa S ( N\bar{N}-M^2) 
\nonumber
\\ 
& & +\lambda_{D}^{i} N D_{i}D_{i}^{c}+ \lambda_{q}^{i}D_{i}qq +
\lambda_{q^c}^{i}D^{c}_{i}u^cd^c\nonumber \\
& & +\lambda_{L}SL\bar{L} +
\lambda_{H_{d}}^{\alpha}\nu^c\bar{L}H_{d}^{\alpha}+ 
\lambda_{N}^{i}N_{i}N_{i}\frac{\bar{N}^{2}}{2m_{\rm P}},
\label{W}
\eea
where $m_{\rm P}$ is the reduced Planck mass and $y_u$, $y_d$, 
$y_{\nu}$, $y_e$ are the Yukawa coupling constants 
with the family indices suppressed. Here 
$q$, $u^c$, $d^c$, $l$, $\nu^c$, $e^c$ are the usual quark and lepton 
superfields of MSSM including the right handed neutrinos $\nu^c$ and 
$H_{u}^{i}$, $H_{d}^{j}$ ($i,j=1,2,3$) are $SU(2)_{\rm L}$ doublets 
with hypercharge $Y=1/2$, $-1/2$ respectively. The superfields $N$, 
$\bar{N}$ constitute a conjugate pair of SM singlets, 
while $S$ is a gauge singlet. The coupling $\lambda_{\mu}^{ij}N
H_{u}^{i}H_{d}^{j}$ is diagonalized by appropriate rotations of 
$H_{u}^{i}$ and $H_{d}^{j}$ and a discrete $Z_2$ symmetry under which 
$H_{u}^{\alpha}$ and $H_{d}^{\alpha}$ ($\alpha=2,3$) are odd is 
imposed. Consequently, only $H_{u}^{1}$, $H_{d}^{1}$ couple to quarks 
and leptons and are the standard electroweak Higgs superfields. 

The superfields $D_i$ and $D_i^c$ ($i=1,2,3$) are color triplets 
and antitriplets with $Y=-1/3$ and $1/3$ respectively and the 
coupling $\lambda_D^{ij}ND_{i}D_{j}^c$ is diagonalized by 
appropriate rotations of $D_i$ and $D^c_j$. The superfields $N_i$ 
($i=1,2,3$) are SM singlets and the coupling $\lambda_{N}^{ij}
N_{i}N_{j}\bar{N}^{2}/2m_{\rm P}$ is again diagonalized by rotating 
$N_{i}$ and $N_{j}$. We impose an extra $Z_2^{\prime}$ symmetry 
under which the $N_i$'s are odd. In order to achieve unification of 
the MSSM gauge coupling constants, we introduced an extra conjugate 
pair of $SU(2)_{\rm L}$ doublets $L$ and $\bar{L}$ with $Y=-1/2$ 
and $1/2$ respectively. These doublets are odd under $Z_2$ and 
together with $H_{d}^{\alpha}$ and $H_{u}^{\alpha}$ ($\alpha=2,3$) 
form three complete $SU(5)$ multiplets with the color 
(anti)triplets $D_i$ and $D_i^c$. Note that the superfields $q$, 
$u^c$, $d^c$, $l$, $\nu^c$, $e^c$, $H_u^{i}$, $H_d^{i}$, $D_i$, 
$D_i^c$, and $N_i$ form three complete fundamental representations 
of $E_6$, while $N$, $\bar{N}$ and $L$, $\bar{L}$ are conjugate 
pairs from incomplete $E_6$ multiplets. 

In Table~\ref{tab:fields}, we summarize all the superfields of 
the model together with their transformation properties under 
the SM gauge group $G_{\rm SM}$ and their charges under the 
discrete symmetries $Z_2$, $Z_2'$, the global R symmetry $U(1)_R$, 
and the local $U(1)_{\psi'}$ with GUT-normalized charge 
$Q_{\psi'}$. Note that the discrete symmetries $Z_2$, $Z_2'$ do 
not carry $SU(3)_{\rm c}$ or $SU(2)_{\rm L}$ anomalies. 

\begin{table}[!t]
\caption{Superfield Content of the Model.}
\begin{tabular}{c@{\hspace{.43cm}}
c@{\hspace{.43cm}} c@{\hspace{.43cm}} c@{\hspace{.43cm}}
c@{\hspace{.43cm}}c}
\toprule
{Superfields}&{Representations}&\multicolumn{4}{c}
{Extra Symmetries}
\\
{}&{under
$G_{\rm SM}$}&{$Z_2$}&{$Z_2'$}&{$R$}&{$2\sqrt{10}Q_{\psi'}$}
\\\colrule
\multicolumn{5}{c}{Matter Superfields}\\\colrule
{$q$} &{$({\bf 3, 2}, 1/6)$}&$+$&$+$& $1/2$& $1$
\\
{$u^c$} & {$({\bf \bar 3, 1},-2/3)$}&$+$&$+$&$1/2$&{$1$}
\\
{$d^c$} & {$({\bf \bar 3, 1},1/3)$}  &$+$&$+$&$1/2$&{$2$}
\\
{$l$} &{$({\bf 1, 2}, -1/2)$} &$+$&$+$&$0$& $2$
 \\
{$\nu^c$} & {$({\bf 1, 1}, 0)$}&{$+$}&$+$ &$1$&{$0$}
\\
{$e^c$} & {$({\bf 1, 1}, 1)$}&{$+$}&$+$&$1$ &{$1$}
\\
{$H_u^{\alpha}$} & {$({\bf 1, 2},1/2)$} &$-$&$+$&$1$&$-2$
\\
{$H_d^{\alpha}$} & {$({\bf 1, 2},-1/2)$} &$-$&$+$&$1$&$-3$
\\
{$D_i$} & {$({\bf 3, 1},-1/3)$} &$+$&$+$ &$1$& $-2$
\\
{$D_i^c$} &{$({\bf \bar 3, 1},1/3)$}&$+$&$+$& $1$&$-3$\\
{$N_i$} &{$({\bf 1, 1},0)$}  &$+$&$-$& $1$&$5$\\
\colrule
\multicolumn{5}{c}{Higgs Superfields}
\\\colrule
{$H_u^{1}$} & {$({\bf 1, 2},1/2)$} &$+$&$+$&$1$&$-2$
\\
{$H_d^{1}$} & {$({\bf 1, 2},-1/2)$} &$+$&$+$&$1$&$-3$
\\
{$S$} & {$({\bf 1, 1},0)$}  &$+$&$+$&$2$ &$0$
\\ 
{$N$} &{$({\bf 1, 1},0)$} & {$+$}&$+$&{$0$}&{$5$} 
\\
{$\bar{N}$}&{$({\bf 1, 1},0)$}&{$+$}&$+$&{$0$}&{$-5$} 
\\
\colrule
\multicolumn{5}{c}{Extra $SU(2)_{\rm L}$ Doublet 
Superfields}
\\\colrule
{$L$} &{$({\bf 1, 2},-1/2)$} & {$-$}&$+$&{$0$}&{$-3$} \\
{$\bar{L}$} &{$({\bf 1, 2},1/2)$} & {$-$}&$+$&{$0$}&{$3$} 
\\
\botrule
\end{tabular}\label{tab:fields}
\end{table}

The symmetries of the model allow
not only the superpotential terms in Eq.~(\ref{W}), but also the 
following higher order terms (divided by appropriate powers of 
$m_{\rm P}$):
\bea
& & \nu^{c}H_{u}^{\alpha}LN, e^cH_d^{\alpha}L\bar{N},
H_{u}^1H_{u}^{1}ll, H_{u}^{\alpha}H_{u}^{\beta}ll,
H_u^1H_d^{\alpha}l\bar{L}, 
\nonumber\\
& & H_u^{\alpha}H_d^1l\bar{L}, H_d^1H_d^1\bar{L}\bar{L},  
H_d^{\alpha}H_d^{\beta}\bar{L}\bar{L}, qu^cqd^c\bar{N}, 
qu^ce^cl\bar{N}, 
\nonumber\\
& & qd^c\nu^cl\bar{N}, e^c\nu^cLLN, H_u^{\alpha}qd^clL, 
H_u^1H_u^{\alpha}lLN, 
\nonumber\\
& & H_u^1H_u^1LLNN, H_u^{\alpha}H_u^{\beta}LLNN, 
H_u^{\alpha}qu^cl\bar{L}\bar{N}, 
\nonumber\\
& &H_d^{\alpha}qd^cl\bar{L}\bar{N}, \nu^cH_d^1l\bar{L}\bar{L}\bar{N}, 
e^cH_u^1lLLN, qd^cLqd^cL, 
\nonumber\\
& & D^c_iu^cu^c\bar{L}\bar{L}\bar{N}, 
D^c_id^cd^cLLN, e^cqd^clLL,  H_u^1qd^cLLN, 
\nonumber\\
& & H_d^1qu^c\bar{L}\bar{L}\bar{N}, 
H_d^{1}H_d^{\alpha}l\bar{L}\bar{L}\bar{L}\bar{N}, 
H_u^{\alpha}e^cLLLNN, 
\nonumber\\
& &  \nu^cqu^cl\bar{L}\bar{L}\bar{N}\bar{N}, 
qu^cqu^c\bar{L}\bar{L}\bar{N}\bar{N}, e^ce^cLLLLNN,
\nonumber\\ 
& &H_d^{\alpha}qu^cl\bar{L}\bar{L}\bar{L}\bar{N}\bar{N}. 
\label{couplings}
\eea
Note that all the couplings in  Eqs.~(\ref{W}) and (\ref{couplings})
can be multiplied by the combinations $N\bar{N}/m_{\rm P}^2$, 
$L\bar{L}/m_{\rm P}^2$, and $\bar{L}l\bar{N}\bar{L}l\bar{N}/
m_{\rm P}^6$ arbitrarily many times and this exhausts all the 
possible superpotential couplings compatible with the symmetries of 
the model. 

Assigning baryon number $B=-2/3$ and $2/3$ to the diquark superfields 
$D_i$ and $D_i^c$, respectively, we see that the baryon number 
$U(1)_B$ symmetry is automatically present to all orders in the 
superpotential and, 
thus, fast proton decay and other baryon number violating effects are 
avoided \cite{Lazarides:1998iq}. 

The fundamental representation of $E_6$ contains two SM singlets with 
the quantum numbers of $\nu^c$ and $N_i$. Let us assume that at high
energies the gauge symmetry is $G_{\rm SM}\times U(1)_{\chi}\times 
U(1)_{\psi}$. A conjugate pair of Higgs superfields of the type 
$\nu^c$, $\bar{\nu}^c$ from an incomplete $E_6$ multiplet can break 
$U(1)_{\chi}\times U(1)_{\psi}$ to $U(1)_{\psi'}$ at a  
scale of order the GUT scale. So, at lower energies, only 
the gauge symmetry $G_{\rm SM}\times U(1)_{\psi'}$ of our model 
survives. The spontaneous breaking of $U(1)_{\psi'}$ at a scale 
$M\sim 10~{\rm TeV}$ is then achieved by a conjugate pair of Higgs 
superfields of the type $N$, $\bar{N}$ from an incomplete $E_6$ 
multiplet via the superpotential terms $\kappa S(N\bar{N}-M^2)$. 
This breaking will generate a network of local 
superconducting strings. Their string tension, which is 
determined by the scale $M$, is relatively small and certainly 
satisfies the most stringent relevant upper bound from pulsar 
timing arrays \cite{pulsar}. Note, in passing, that the kinetic 
mixing of $U(1)_{\psi'}$ and $U(1)_Y$ is negligible -- see 
fourth paper in Ref.~\cite{chipsi}.      

The `bare' MSSM $\mu$ term is replaced by a term $\lambda_{\mu}^{1}
NH_{u}^1H_{d}^1$, so that the $\mu$ term is generated after $N$ 
acquires a non-zero vacuum expectation value (VEV) $\vev{N}$ of 
order $10~{\rm TeV}$. The same VEV gives 
masses to the two remaining pairs of $SU(2)_{\rm L}$ doublets 
$H_{u}^{\alpha}$, $H_{d}^{\alpha}$ ($\alpha=2,3$) via the 
superpotential terms $\lambda_{\mu}^{\alpha}NH_{u}^{\alpha}
H_{d}^{\alpha}$ as well as to the diquarks $D_i$, $D_i^c$ 
($i=1,2,3$) via the terms  $\lambda_{D}^{i}ND_iD_i^c$.
The gauge singlet $S$ acquires a VEV $\vev{S}$ of order 
TeV from soft SUSY 
breaking \cite{Dvali:1997uq}. (In the SUSY limit the VEV of 
$S$ is zero.) This VEV generates masses for the extra doublets
$L$, $\bar{L}$ via the term $\lambda_{L}SL\bar{L}$. Finally, 
the sterile neutrino fields, which are the fermionic parts of 
$N_i$, acquire masses of order $10^{-1}~{\rm eV}$ or so via the 
terms $\lambda_{N}^{i}N_iN_i\bar{N}^2/2m_{\rm P}$. 

The spontaneous breaking of $U(1)_{\psi'}$ implemented with the 
fields $S$, $N$, $\bar{N}$ delivers, in the exact SUSY limit, 
four spin zero particles all with the same mass given by 
$\sqrt{2}\kappa M$. This mass, even for $M\gg 1~{\rm TeV}$, can 
be of order TeV by selecting an appropriate value for $\kappa$. 
We should point out though that, depending on the SUSY breaking 
mechanism, these states may end up with significantly different 
masses. The diquarks $D_i$, $D_i^c$ may be found \cite{diquark} 
at the LHC.

\section{$U(1)_{\psi'}$ breaking}
\label{sec:details}

We will assume here that the breaking scale of $U(1)_{\psi'}$
is much bigger than the electroweak scale. In this case, the 
spontaneous breaking of $U(1)_{\psi'}$ is not affected by the 
electroweak Higgs doublets in any essential way and can be 
discussed by considering only the superpotential terms 
\beq
\delta W=\kappa S(N\bar{N}-M^2)
\eeq 
in the right-hand side (RHS) of Eq.~(\ref{W}). They give the 
following scalar potential
\bea
V&=&\kappa^2|N\bar{N}-M^2|^2+ 
\kappa^2|S|^2(|N|^2+|\bar{N}|^2)
\nonumber\\
& &+\left(A\kappa S N\bar{N}-(A-2m_{3/2})\kappa M^2 S+
{\rm H.c.}\right)
\nonumber\\
& &+m_{0}^2(|N|^2+|\bar{N}|^2+|S|^2)+{\rm D-terms}.
\label{V}
\eea
Here the mass parameter $M$ and the dimensionless coupling 
constant $\kappa$ are made real and positive by field 
rephasing and the scalar components of the superfields are 
denoted by the same symbol. The parameter $m_{3/2}$ is the 
gravitino mass, $A\sim m_{3/2}$ is the coefficient of the 
trilinear soft terms taken real and positive, and $m_0\sim 
m_{3/2}$ is the common soft mass of $N$, $\bar{N}$, and 
$S$. We assumed, 
for definiteness, minimal supergravity. In this case, the 
coefficients of the trilinear and linear soft terms are 
related as shown in Eq.~(\ref{V}). Vanishing of the 
D-terms implies that 
$|N|=|\bar{N}|$, which yields $\bar{N}^*=e^{i\vartheta}N$, 
while minimization of the potential requires that 
$\vartheta=0$. So, $N$ and $\bar{N}$ can be rotated to the 
positive real axis by a $U(1)_{\psi'}$ transformation.
 
We find \cite{Dvali:1997uq} that the scalar potential in 
Eq.~(\ref{V}) is mini\-mized at
\beq
\vev{S}=-\frac{m_{3/2}}{\kappa}\left(1+\sum_{n\geq 1}
c_n\left(\frac{m_{3/2}}{M}\right)^n\right)
\eeq
and 
\beq
\vev{N}=\vev{\bar{N}}\equiv\frac{N_0}{\sqrt{2}}=
M\left(1+\sum_{n\geq 1}
d_n\left(\frac{m_{3/2}}{M}\right)^n\right),
\eeq
where $c_n$, $d_n$ are numerical coefficients of order 
unity. Assuming that $M\gg m_{3/2}$ and keeping in 
$\vev{S}^2$ and $N_0^2$ terms up to order $m_{3/2}^2$, 
these formulas can be approximated as follows:
\beq
\vev{S}\simeq-\frac{m_{3/2}}{\kappa}, \quad 
\frac{N_0^2}{2}\simeq M^2+\frac{Am_{3/2}-
m_{3/2}^2-m_0^2}{\kappa^2}.
\label{VEVs}
\eeq
We should point out that the trilinear and linear soft 
terms in the second line of Eq.~(\ref{V}) play an important 
role in our scheme. Substituting $N$ and $\bar{N}$ by their 
VEVs, these terms yield a linear term in $S$ which, together 
with the mass term of $S$, generates \cite{Dvali:1997uq} a 
VEV for $S$ of order TeV. It is then obvious that, 
substituting this VEV of $S$ in the superpotential term 
$\lambda_L S L\bar{L}$, the superfields $L$, $\bar{L}$ 
acquire a mass $m_L=\lambda_L|\vev{S}|=\lambda_L m_{3/2}/
\kappa$. Moreover, the MSSM $\mu$ term is obtained by 
substituting $\vev{N}$ in the superpotential term 
$\lambda_{\mu}^{1}NH_u^{1}H_d^{1}$ with $\mu=
\lambda_{\mu}^{1}N_0/\sqrt{2}$, while $H_u^{\alpha}$, 
$H_d^{\alpha}$ ($\alpha=2,3)$ and $D_i$, $D_i^c$ 
acquire masses of order TeV from the 
couplings $\lambda_{\mu}^{\alpha}NH_u^{\alpha}H_d^{\alpha}$
and $\lambda_{D}^iND_iD_i^c$ respectively. Note that, with 
$D_i$, $D_i^c$, $L$, $\bar{L}$, and $H_u^{\alpha}$, 
$H_d^{\alpha}$ masses $\sim {\rm TeV}$, the gauge couplings 
stay in the perturbative domain for up to four such pairs
of color (anti)triplets and $SU(2)_{\rm L}$ doublets. 

The mass spectrum of the scalar $S-N-\bar{N}$ system can 
be constructed by substituting $N=\vev{N}+\delta\tilde{N}$ and 
$\bar{N}=\vev{\bar{N}}+\delta\tilde{\bar{N}}$. In the 
unbroken SUSY limit, we find two complex scalar fields 
$S$ and $\theta=(\delta\tilde{N}+\delta\tilde{\bar{N}})/
\sqrt{2}$ with equal masses $m_{S}=m_{\theta}=\sqrt{2}
\kappa M$. Soft SUSY breaking can, of 
course, mix these fields and generate a mass splitting. For 
example, the trilinear soft term $A\kappa S N\bar{N}$ yields 
a mass-squared splitting $\pm\sqrt{2}\kappa M A$ with the 
mass eigenstates now being $(S+\theta^*)/\sqrt{2}$ and 
$(S-\theta^*)/\sqrt{2}$. This splitting is small for 
$A\ll\sqrt{2}\kappa M$.

\section{Electroweak Symmetry Breaking}
\label{sec:EWSB} 

The standard scalar potential for the radiative 
electroweak symmetry breaking in MSSM is modified in the 
present model. A modification originates from the 
D-term for $U(1)_{\psi'}$: 
\beq
V_D=\frac{g_{\psi'}^2}{80}\left[-2|H_u|^2-3|H_d|^2+
5\left(|N|^2-|\bar{N}|^2\right)\right]^2,
\label{VD0}
\eeq 
where $g_{\psi'}$ is the GUT-normalized gauge coupling 
constant for the $U(1)_{\psi'}$ symmetry and $H_u$, $H_d$ 
are the neutral components of the scalar parts of the 
Higgs $SU(2)_{\rm L}$ doublet superfields $H_u^1$, $H_d^1$ 
respectively. In order to find the leading contribution of 
this D-term to the electroweak potential, we must 
integrate out to one loop the heavy degrees of freedom $N$ 
and $\bar{N}$. To this end, we express these complex scalar 
fields in terms of the canonically normalized real scalar 
fields $\delta N$, $\delta\bar{N}$, $\varphi$, 
$\bar{\varphi}$ as follows:
\beq
N=\frac{1}{\sqrt{2}}(N_0+\delta N)e^{\frac{i\varphi}{N_0}},
\quad \bar{N}=\frac{1}{\sqrt{2}}(N_0+\delta\bar{N})
e^{\frac{i\bar{\varphi}}{N_0}}.
\label{etatheta}
\eeq
Then the combination $|N|^2-|\bar{N}|^2$, which appears in 
the D-term in Eq.~(\ref{VD0}), becomes
\beq
|N|^2-|\bar{N}|^2=\sqrt{2}N_0\eta+\eta\xi,
\eeq
where 
\beq
\eta=\frac{\delta N-\delta\bar{N}}{\sqrt{2}}, \quad
\xi=\frac{\delta N+\delta\bar{N}}{\sqrt{2}}
\eeq 
are canonically normalized real scalar fields. The D-term
can now be expanded as follows:
\beq
V_D=\frac{g_{\psi'}^2}{80}\left[E^2+10\sqrt{2}N_0E\eta+
50N_0^2\eta^2+\cdots \right],
\label{VD}
\eeq
where $E\equiv-2|H_u|^2-3|H_d|^2$. Here we kept only up
to quadratic terms in $\eta$, $\xi$, but ignored the
mixed quadratic term proportional to $\eta\xi$ since
its coefficient is much smaller than the coefficient of 
the $\eta^2$ term assuming that $N_0$ is much bigger 
than the electroweak scale.

We see, from Eq.~(\ref{VD}), that integrating out the 
heavy states reduces to the calculation of a path integral 
over the real scalar field $\eta$. To do this, we first 
need to find the $\eta$ dependence of the potential $V$
in Eq.~(\ref{V}). So we substitute in this equation $N$ 
and $\bar{N}$ from Eq.~(\ref{etatheta}). Keeping only 
$\eta$-dependent terms up to the second order and 
substituting $S$ by its VEV in Eq.~(\ref{VEVs}), we 
obtain
\beq
\delta V\simeq \frac{1}{2}\left(-\frac{\kappa^2}{2}N_0^2+
m_{3/2}^2+m_0^2+\kappa^2M^2+Am_{3/2}\right)\eta^2,
\eeq
which, substituting $N_0$ from Eq.~(\ref{VEVs}), gives
\beq
\delta V\simeq m_N^2\eta^2 \quad {\rm with} \quad m_N^2\equiv 
m_{3/2}^2+m_0^2.
\eeq   
Adding $\delta V$ to the D-term potential in 
Eq.~(\ref{VD}), we obtain the potential
\beq
V_{\eta}=\frac{g_{\psi'}^2}{80}E^2
+\frac{\sqrt{2}g_{\psi'}^2}{8}N_0E\eta+\left(m_N^2+
\frac{5g_{\psi'}^2}{8}N_0^2\right)\eta^2+\cdots,
\eeq 
which can be given the form
\bea
V_{\eta}&=&\frac{g_{\psi'}^2E^2}{80}\left(1+
\frac{5g_{\psi'}^2N_0^2}{8m_N^2}\right)^{-1}
+\left(m_N^2+\frac{5g_{\psi'}^2N_0^2}{8}\right)
\nonumber\\
& & \times\left(\eta+\frac{g_{\psi'}^2N_0E}{8\sqrt{2}
\left(m_N^2+\frac{5g_{\psi'}^2N_0^2}{8}\right)}
\right)^2+\cdots.
\eea
The path integral 
\beq
\int (d\eta)e^{-iV_{\eta}\cal{V}},
\eeq
where $\cal{V}$ is the spacetime volume, can be readily
calculated and, besides an irrelevant overall constant 
factor, we are left with the term 
\beq
\delta V_D\simeq\frac{g_{\psi'}^2}{80}\left[2|H_u|^2
+3|H_d|^2
\right]^2\left(1+\frac{m_{Z'}^2}{2m_N^2}\right)^{-1}
\label{VDfinal}
\eeq 
to be added to the usual electroweak symmetry breaking 
potential. Here $m_{Z'}=\sqrt{5}g_{\psi'}N_0/2$ is the 
mass of the $Z'$ gauge boson associated with 
$U(1)_{\psi'}$.

Another modification of the MSSM electroweak potential 
comes from the integration of the heavy complex field 
$S$ with mass $\sqrt{2}\kappa M$ in the exact SUSY 
limit. The cross F-term $F_N$ between the superpotential 
terms $\kappa SN\bar{N}$ and $\lambda_\mu^1NH_u^1H_d^1$ 
in Eq.~(\ref{W}) together with the mass-squared term of 
$S$ give
\bea
& &2\kappa^2M^2|S|^2+\left(\kappa S^*
\bar{N}^*\tilde{\lambda}_\mu H_u^1H_d^1+{\rm H.c.}
\right)= 
\nonumber\\
& &\left|\sqrt{2}\kappa MS+\frac{1}{\sqrt{2}}
\tilde{\lambda}_\mu H_u^1H_d^1\right|^2-\frac{1}{2}
\tilde{\lambda}_\mu^2
|H_u^1H_d^1|^2,
\eea
where $\tilde{\lambda}_\mu\equiv\lambda_\mu^1$.
Integrating out $S$, we then obtain the extra term
\beq
-\frac{1}{2}\tilde{\lambda}_\mu^2|H_u|^2|H_d|^2
\label{lambdaterm}
\eeq
in the electroweak potential. One can show that the 
integration of all the other heavy fields gives 
smaller contributions, which we ignore.

Now the potential for the electroweak symmetry 
breaking as can be derived from the superpotential 
terms
\beq
\kappa S(N\bar{N}-M^2)-\tilde{\lambda}_\mu NH_uH_d
\label{WEW}
\eeq 
after substituting the VEVs of $S$, $N$, and $\bar{N}$
from Eq.~(\ref{VEVs}) and adding the D-term in 
Eq.~(\ref{VDfinal}) and the term in Eq.~(\ref{lambdaterm}) is
\bea
V_{\rm EW} &\simeq & m^2_{H_u} |H_u|^2 + m^2_{H_d} |H_d|^2-B 
(H_u H_d+{\rm H.c.})
\nonumber\\
& & +\lambda_\mu^2|H_u|^2|H_d|^2+
\frac{1}{8}(g^2 + {g^\prime}^2)(|H_u|^2-|H_d|^2)^2 
\nonumber\\
& &+ c(Q_u|H_u|^2 + Q_d|H_d|^2)^2,
\label{VEW}
\eea
where $m^2_{H_u}=\tilde{m}^2_{H_u}+\mu^2$,  
$m^2_{H_d}=\tilde{m}^2_{H_d}+\mu^2$ with $\tilde{m}_{H_u}$,
$\tilde{m}_{H_d}$ being the soft masses of $H_u$, $H_d$ and 
$B=\tilde{B}-m_{3/2}$ with $\tilde{B}$ being the coefficient 
of the soft trilinear term corresponding to the second term 
in Eq.~(\ref{WEW}). Here $\lambda_{\mu}\equiv
\tilde{\lambda}_{\mu}/\sqrt{2}$, $g$ is the $SU(2)_{\rm L}$ 
and $g'$ the non-GUT-normalized $U(1)_Y$ gauge coupling 
constant, $Q_u=2$, $Q_d=3$, and
\beq
c=\frac{g_{\psi'}^2}{80}\left(1+\frac{m_{Z'}^2}{2m_N^2}
\right)^{-1}.
\eeq
Note that the potential in Eq.~(\ref{VEW}) contains the 
so-called next-to-minimal supersymmetric standard model 
(NMSSM) term
\beq
\lambda_\mu^2|H_u|^2|H_d|^2.
\eeq

Minimization of the potential in Eq.~(\ref{VEW}) yields the 
following relations:
\begin{eqnarray}
m^2_{H_u} &=& m_A^2 \cos^2\beta +\frac{1}{2}m_Z^2\cos 2\beta-
\lambda_\mu^2v^2\cos^2\beta\nonumber\\
& & -2cQ_uv^2(Q_u\sin^2\beta+Q_d\cos^2\beta),\nonumber\\
m^2_{H_d} &=& m_A^2\sin^2\beta- \frac{1}{2}m_Z^2\cos 2\beta- 
\lambda_\mu^2v^2\sin^2\beta\nonumber\\
& &-2cQ_dv^2(Q_u\sin^2\beta+Q_d\cos^2\beta).
\label{min}
\end{eqnarray}
Here $v^2=v_u^2+v_d^2$ with $v_u=\vev{H_u}$ and $v_d=\vev{H_d}$, 
$\tan\beta=v_u/v_d$, and the expressions 
\beq
m_Z^2=\frac{1}{2}(g^2+g'^2)v^2, \quad m_A^2=\frac{2B\mu}
{\sin 2\beta}
\eeq
for the $Z$ gauge boson mass $m_Z$ and the CP-odd Higgs boson 
mass $m_A$ are used. Note that the latter is not affected by 
the extra terms in the potential $V_{\rm EW}$ since they 
involve only the absolute values of $H_u$, $H_d$. 

The mass-squared matrix in the CP-even Higgs sector 
\begin{equation}
\mathcal{M} = 
\begin{pmatrix} 
\mathcal{M}_{11} & \mathcal{M}_{12}\\
& \\
\mathcal{M}_{12} & \mathcal{M}_{22}
\end{pmatrix}
\label{matrix}
\end{equation}
can be constructed by substituting $H_u=v_u+h_u/\sqrt{2}$
and $H_d=v_d+h_d/\sqrt{2}$ in the RHS of Eq.~(\ref{VEW}) and 
keeping only terms quadratic in $h_u$, $h_d$. We find
\begin{eqnarray}
\mathcal{M}_{11}&=& m^2_{H_u}+\frac{1}{2}m_Z^2(3\sin^2\beta-\cos^2\beta)
+\lambda_\mu^2v^2\cos^2\beta
\nonumber\\
& &+2cQ_uv^2\left(3Q_u\sin^2\beta+Q_d\cos^2\beta\right),
\nonumber\\
\mathcal{M}_{12}&=&\left(-m_A^2-m_Z^2+
2\lambda_\mu^2v^2+4cQ_uQ_dv^2\right)\sin\beta\cos\beta, 
\nonumber\\
\mathcal{M}_{22} &=&m^2_{H_d}+\frac{1}{2}m_Z^2 (3\cos^2\beta-2\sin^2\beta)
+\lambda_\mu^2v^2\sin^2\beta
\nonumber\\
& & + 2cQ_dv^2(3Q_d\cos^2\beta+Q_u\sin^2\beta).
\end{eqnarray}
Using the minimization conditions in Eq.~(\ref{min}), $\mathcal{M}_{11}$
and  $\mathcal{M}_{22}$ can be cast in the form
\begin{eqnarray}
\mathcal{M}_{11}&=& m_A^2 \cos^2\beta+(m_Z^2+4cQ_u^2v^2)\sin^2\beta,
\nonumber\\
\mathcal{M}_{22} &=& m_A^2\sin^2\beta+(m_Z^2+4cQ_d^2v^2)\cos^2\beta.
\end{eqnarray}
The eigenvalues $m_h^2$ and $m_H^2$ of the mass-squared matrix in 
Eq.~(\ref{matrix}), which are, respectively, the `tree-level' 
masses squared of the lightest and heavier neutral CP-even Higgs 
bosons, can now be constructed: 
\beq
m_{h,H}^2=\frac{1}{2}\Sigma\mp\sqrt{\frac{1}{4}\Sigma^2-\Delta}
\eeq
with
\bea
\Sigma&=&m_A^2+m_Z^2+4c^2v^2(Q_u^2\sin^2\beta+Q_d^2
\cos^2\beta),
\nonumber\\
\Delta&=&m_A^2m_Z^2\cos^2 2\beta+
4cv^2m_A^2(Q_u\sin^2\beta+Q_d\cos^2\beta)^2
\nonumber\\
&&+\lambda_\mu^2v^2m_A^2\sin^2 2\beta+cv^2m_Z^2
(Q_u+Q_d)^2\sin^2 2\beta
\nonumber\\
&&+m_Z^2\lambda_\mu^2 v^2\sin^2 2\beta-\lambda_\mu^4 v^4
\sin^2 2\beta
\nonumber\\
&&-4cQ_uQ_d\lambda_\mu^2 v^4\sin^2 2\beta.
\eea
Let us note that, here, by `tree-level' masses we mean 
the masses without the inclusion of the radiative 
corrections in MSSM. It is easy to see that $m_h^2$, 
in the so-called decoupling limit where $m_A\gg m_Z$, 
is given by
\bea
m_h^2&=&m_Z^2\cos^2 2\beta+4cv^2(Q_u\sin^2\beta+
Q_d\cos^2\beta)^2
\nonumber\\
&&+\lambda_\mu^2v^2\sin^2 2\beta.
\label{mh}
\eea
 
\section{Diphoton Resonances}
\label{sec:diphoton}

\begin{figure}[t]
\centerline{\epsfig{file=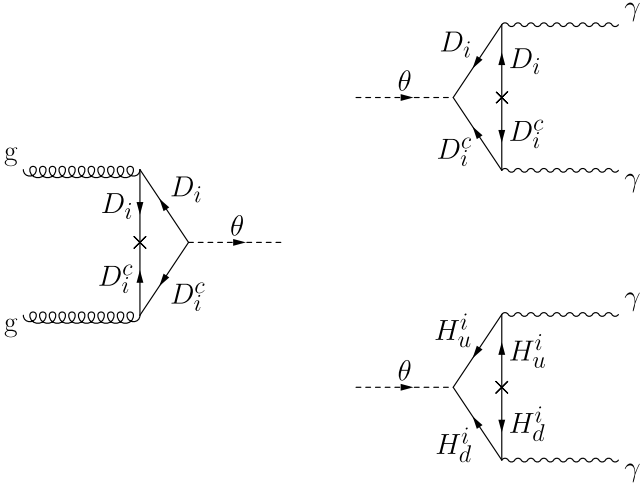,width=8.7cm}}
\caption{Production of the complex scalar field $\theta$ at 
the LHC by gluon ($g$) 
fusion and its subsequent decay into photons ($\gamma$). Solid 
(dashed) lines represent the fermionic (bosonic) component of 
the indicated superfields. The arrows depict the chirality of 
the superfields and the crosses are mass insertions which must 
be inserted in each of the lines in the loops.}
\label{fig1}
\end{figure}

The real scalar $\theta_1$ and real pseudoscalar $\theta_2$ 
components of $\theta=(\delta\tilde{N}+\delta\tilde{\bar{N}})
/\sqrt{2}~[=(\theta_1+i\theta_2)/\sqrt{2}\,]$ with mass 
$m_\theta=\sqrt{2}\kappa M$ in the exact SUSY limit can be 
produced at the LHC by gluon fusion via a fermionic $D_i$, 
$D^c_i$ loop as indicated in Fig.~\ref{fig1}. They can decay 
into gluons, photons, $Z$ or $W^{\pm}$ gauge bosons via 
the same loop diagram as well as a similar fermionic 
$H_u^i$, $H_d^i$ loop. The most promising decay channel to 
search for these resonances is into two photons with the 
relevant diagrams also shown in Fig.~\ref{fig1}. 

Applying the results of Ref.~\cite{king}, the cross section 
of the diphoton excess is 
\beq
\label{sigmaK}
\sigma(pp\rightarrow \theta_m\rightarrow\gamma\gamma)\simeq 
\frac{C_{gg}}{m_{\theta}s\Gamma_{\theta_m}}
\Gamma(\theta_m\rightarrow gg)\Gamma(\theta_m\rightarrow 
\gamma\gamma),
\eeq     
where $m=1,2$, $C_{gg}\simeq 3163$, $\sqrt{s}\simeq 
13~{\rm TeV}$, $\Gamma_{\theta_m}$ is the total decay width 
of $\theta_m$, and the decay widths of $\theta_m$ to two 
gluons ($g$) or two photons ($\gamma$) are given by 
\bea
\Gamma(\theta_m\rightarrow gg)&=&\frac{m_\theta^3\alpha_s^2}
{512\,\pi^3\vev{N}^2}\, \left(\sum_{i=1}^{3}A_m(x_i)\right)^2,\\
\Gamma(\theta_m\rightarrow \gamma\gamma)&=&\frac{m_\theta^3
\alpha_Y^2\cos^4\theta_W}
{9216\,\pi^3\vev{N}^2}\left[\sum_{i=1}^{3}A_m(x_i)+\right.
\nonumber\\
& &\left.\frac{3}{2}\sum_{i=1}^{3}A_m(y_i)
\left(1+\frac{\alpha_2\tan^2\theta_W}
{\alpha_Y}\right)\right]^2.
\label{SgammaL}
\nonumber\\
\eea
Here $A_1(x)=2x[1+(1-x)\arcsin^2(1/\sqrt{x})]$, 
$A_2(x)=2x\arcsin^2(1/\sqrt{x})$, $x_i=4m_{D_i}^2/m_\theta^2> 1$ 
with $m_{D_i}=\lambda_{D}^i\vev{N}$ being the mass of $D_i$ and 
$D_i^c$, $y_i=4m_{H_i}^2/m_\theta^2> 1$ with $m_{H_i}=
\lambda_{\mu}^i\vev{N}$ being the mass of $H_u^i$ and 
$H_d^i$, and $\alpha_s$, 
$\alpha_Y$, and $\alpha_2$ are the strong, hypercharge, 
and $SU(2)_{\rm L}$ fine-structure constants, respectively. 

The cross section in Eq.~(\ref{sigmaK}) simplifies under the 
assumption that the spin zero fields $\theta_m$ decay 
predominantly into gluons, namely $\Gamma_{\theta_m}\simeq
\Gamma(\theta_m\rightarrow gg)$. In this case, one obtains 
\cite{franceschini}
\beq
\sigma(pp\rightarrow \theta_m\rightarrow\gamma\gamma)\simeq 
7.3\times 10^{6}\,~\frac{\Gamma(\theta_m\rightarrow\gamma
\gamma)}{m_{\theta}}\,~{\rm fb}.
\label{sigmaF}
\eeq
For $x_i$ and $y_i$ just above unity, which guarantees that 
the decay of $\theta_m$ to $D_i$, $D^c_i$ and $H_u^i$, 
$H_d^i$ pairs is kinematically blocked, $A_1(x_i)$ and 
$A_2(y_i)$ are maximized with values $A_1\simeq 2$ and 
$A_2\simeq \pi^2/2$. So we consider this case. It is also 
more beneficial to consider the decay of the pseudoscalar 
$\theta_2$ since $A_2(x)>A_1(x)$ for all $x>1$. Using 
Eq.~(\ref{SgammaL}), we then find that Eq.~(\ref{sigmaF})
gives
\beq
\sigma(pp\rightarrow \theta_2\rightarrow\gamma\gamma)
\simeq 5.5\left(\frac{m_{\theta}}{\vev{N}}\right)^2
{\rm fb}\simeq 11\,\kappa^2~{\rm fb}.
\label{fb}
\eeq

\begin{figure}[t]
\centerline{\epsfig{file=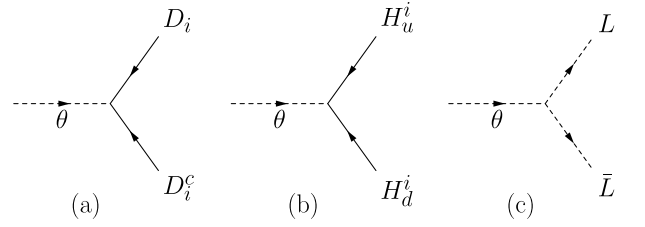,width=8.2cm}}
\caption{Decay of the complex scalar field $\theta$ into a 
fermionic $D_i$, $D_i^c$ (a) or $H_u^i$, $H_d^i$ (b) pair or 
a bosonic $L$, $\bar{L}$ pair (c). The notation is the 
same as in Fig.~\ref{fig1}.}
\label{fig2}
\end{figure}

\begin{figure}[t]
\centerline{\epsfig{file=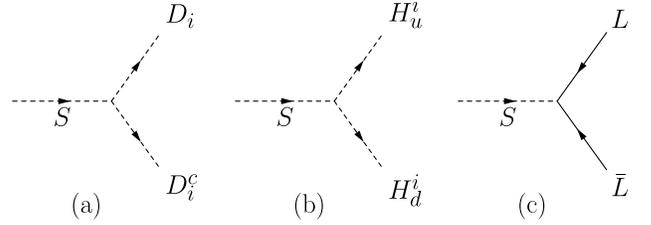,width=8.4cm}}
\caption{Decay of the complex scalar field $S$ into a 
bosonic $D_i$, $D_i^c$ (a) or $H_u^i$, $H_d^i$ (b) pair 
or a fermionic $L$, $\bar{L}$ pair (c). The notation is 
the same as in Fig.~\ref{fig1}.}
\label{fig3}
\end{figure}

In the exact SUSY limit, the complex scalar field $\theta$ could 
decay into a fermionic $D_i$, $D^c_i$ or $H_u^i$, $H_d^i$ pair 
via the superpotential terms $\lambda_{D}^iND_iD_i^c$ or 
$\lambda_{\mu}^iNH_{u}^iH_{d}^i$ if this is kinematically allowed 
-- see Figs.~\ref{fig2}(a) and \ref{fig2}(b). It could also decay 
into a bosonic $L$, $\bar{L}$ pair via the F-term $F_S$ between 
the superpotential couplings $\kappa SN\bar{N}$ and 
$\lambda_L SL\bar{L}$ if this is kinematically allowed -- see 
Fig.~\ref{fig2}(c). The decay widths in the three cases are
\beq
\Gamma^\theta_{D^i}=\frac{(\lambda_{D}^i)^2}{16\pi}m_{\theta}, 
\quad 
\Gamma^\theta_{H^i}=\frac{(\lambda_{\mu}^i)^2}{16\pi}m_{\theta},
\quad
\Gamma^\theta_{L}=\frac{(\lambda_{L})^2}{8\pi}m_{\theta},
\label{decay}
\eeq 
respectively, where we assumed that the mass of the re\-levant
$D_i$, $D^c_i$, or $H_u^i$, $H_d^i$, or $L$, $\bar{L}$ is 
much smaller than $m_{\theta}/2$. Depending on the kinematics 
the total decay width of the resonance could easily lie in the 
$100~{\rm GeV}$ range. The diphoton, dijet, and diboson decay 
modes in this case would be subdominant. 

Our estimate in Eq.~(\ref{sigmaF}) holds provided that the decay 
widths of $\theta$ into a $D_i$, $D^c_i$, or $H_u^i$, $H_d^i$,
or $L$, $\bar{L}$ pair are sub-dominant or these decays are 
kinematically blocked. The latter is achieved for 
$m_\theta\simeq\sqrt{2}\kappa M<2m_{D_i}\simeq 2
\lambda_{D}^i M$, $2m_{H_i}\simeq 2\lambda_{\mu}^i M$, 
and $2m_L\simeq 2\lambda_L |\vev{S}|\simeq 2\lambda_L m_{3/2}/
\kappa$, which implies that 
\beq
\kappa\lesssim \sqrt{2}\lambda_{D}^i,\, 
\sqrt{2}\lambda_{\mu}^i, \,
2\lambda_L \frac{m_{3/2}}{m_\theta}.
\label{ineq}
\eeq
Note that the estimate of the maximal cross section of the 
diphoton excess in Eq.~(\ref{fb}) corresponds to saturating 
the first two of the inequalities in Eq.~(\ref{ineq}). For
simplicity and for not disturbing the MSSM gauge coupling 
unification, we choose to saturate the third inequality 
too. 
 
The complex scalar field $S$ can decay into a bosonic $D_i$, 
$D^c_i$ or $H_u^i$, $H_d^i$ pair via the F-terms $F_N$ 
between the superpotential couplings $\kappa S N\bar{N}$ 
and $\lambda_{D}^iND_iD_i^c$ or $\lambda_{\mu}^iNH_{u}^i
H_{d}^i$ if this is kinematically allowed -- see 
Figs.~\ref{fig3}(a) and \ref{fig3}(b).  It could also decay 
into a fermionic $L$, $\bar{L}$ pair via the superpotential 
coupling $\lambda_L SL\bar{L}$ if this is kinematically 
allowed -- see Fig.~\ref{fig3}(c). The decay widths 
$\Gamma^S_{D^i}$, $\Gamma^S_{H^i}$, and $\Gamma^S_{L}$ in 
the three cases are, respectively, equal to the decay 
widths $\Gamma^\theta_{D^i}$, $\Gamma^\theta_{H^i}$, and 
$\Gamma^\theta_{L}$ in Eq.~(\ref{decay}). It is obvious 
that, if the inequalities in Eq.~(\ref{ineq}) are satisfied 
so as our estimate of the cross section of the diphoton 
excess in Eq.~(\ref{fb}) to hold, these decay channels of 
$S$ are also blocked. In this case, $S$ will decay to 
lighter particles.

Note that, in the exact SUSY limit, the complex scalar field 
$S$ cannot be produced at the LHC by gluon fusion and, thus, 
cannot lead to diphoton excess. This would require bosonic 
$D_i$, $D_i^c$ loops with mass-squared insertions originating 
from soft trilinear SUSY breaking terms -- for such loops see 
Ref.~\cite{diphoton}. As we already mentioned, the soft SUSY 
breaking terms generate mixing between the scalar fields $S$ 
and $\theta$. Consequently, we can have four diphoton 
resonance states rather than just two from the scalar 
$\theta$ alone. Soft SUSY breaking also gives rise to more 
diagrams contributing to the diphoton excess. However,
our estimate of the cross section of the diphoton excess 
for exact SUSY is the dominant one provided that the scale
of $U(1)_{\psi'}$ breaking is much bigger than the soft 
SUSY breaking scale. Finally, let us note that demanding 
that the mass of the $Z'$ gauge boson $m_{Z'}\simeq\sqrt{5}
g_{\psi'}M/\sqrt{2}>3.8~{\rm TeV}$ \cite{Zprime}, say, 
we find that 
\beq
g_{\psi'}M\gtrsim 2.4~{\rm TeV}.
\label{gpsi'M} 
\eeq

\begin{figure}[t]
\centerline{\epsfig{file=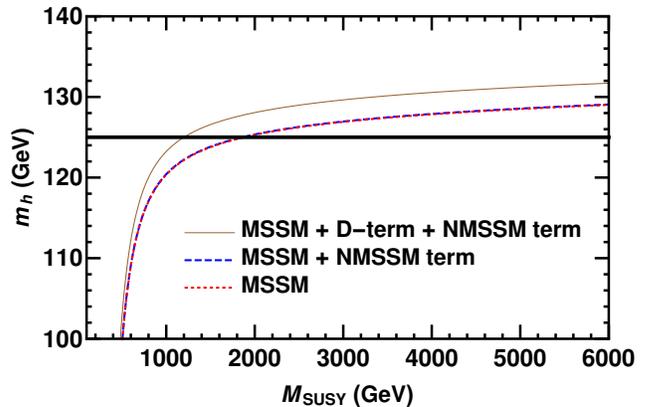,width=8.4cm}}
\caption{Higgs boson mass $m_h$ in the decoupling limit 
and for maximal stop quark mixing versus $M_{\rm SUSY}$ 
for $M=10~{\rm TeV}$, $\tilde{\lambda}_{\mu}=0.3$, 
$\tan\beta=20$, and $m_{3/2}=4~{\rm TeV}$. The dotted 
(red) curve corresponds to MSSM, the dashed (blue) curve
to MSSM plus the NMSSM correction, and the conti\-nuous 
(brown) curve to MSSM plus the D-term and NMSSM 
corrections. The experimental value of $m_h$ is also
depicted by the bold horizontal line.}
\label{higgsmsusy}
\end{figure}

\section{Numerical Analysis}
\label{sec:analysis}

We can show that the gauge coupling constant $g_{\psi'}$
associated with the $U(1)_{\psi'}$ gauge symmetry unifies 
with the MSSM gauge coupling constants provided that its 
value at low energies is equal to about 0.45. 
This value depends very little on the exact value of the 
diquark, the extra $SU(2)_{\rm L}$ doublet, the 
resonance, and the $Z'$ gauge supermultiplet masses. 
So the bound in Eq.~(\ref{gpsi'M}) implies that 
$M\gtrsim 5.34~{\rm TeV}$. As an example, we will set 
$M=10~{\rm TeV}$. In addition, we can show that the 
coupling constants $\kappa$ and $\tilde{\lambda}_{\mu}$
remain perturbative up to the GUT scale provided that
they are not much bigger than about 0.7. 
The requirement that the diphoton resonance mass 
$m_\theta=\sqrt{2}\kappa M$ is bigger than about 
$4.5~{\rm TeV}$ as indicated by the recent CMS results 
\cite{cmsdiphoton}, implies that $\kappa\gtrsim 0.32$. 
In the case where the first two inequalities in 
Eq.~(\ref{ineq}) are saturated, we then obtain that 
$0.5\gtrsim\lambda_{D}^i, \lambda_{\mu}^i\gtrsim 0.22$.
For definiteness, we choose $\lambda_{D}^i\simeq
\lambda_{\mu}^i\simeq 0.3$, which means in particular 
that $\tilde{\lambda}_{\mu}\simeq 0.3$. This choice 
implies that $\kappa\simeq 0.42$, $m_{D_i}\simeq 
m_{H_i}\simeq 3~{\rm TeV}$ (in particular $\mu\simeq 
3~{\rm TeV}$), $m_{\theta}\simeq 6~{\rm TeV}$, and 
$m_{Z'}\simeq 7.1~{\rm TeV}$. Saturating the third 
inequality in Eq.~(\ref{ineq}), we obtain $m_L\simeq 
3~{\rm TeV}$.  Note that, for $\kappa\lesssim 0.7$, 
the resonance mass remains below $9.9~{\rm TeV}$. 

In Fig.~\ref{higgsmsusy}, we plot the lightest CP-even 
Higgs boson mass $m_{h}$ in the decoupling limit versus 
$M_{\rm SUSY}$, which is the geometric mean of the stop 
quark mass eigenva\-lues. We generally assume maximal stop 
quark mixing, which maximizes $m_h$, and include the 
two-loop radiative corrections to $m_h$ in MSSM using 
the package SUSYHD \cite{susyhd}. The NMSSM and D-term 
contributions 
to $m_h$ are also included from Eq.~(\ref{mh}). In this 
figure, $\tan\beta=20$ and $m_{3/2}=4~{\rm TeV}$. 
Notice that the NMSSM correction is very small since 
$\tilde{\lambda}_{\mu}$ is relatively small. The D-term 
correction, however, is sizable and allows us to obtain 
the observed value of $m_h$ with much smaller stop quark 
masses than the ones required in MSSM or NMSSM. Indeed, 
the inclusion of the D-term from $U(1)_{\psi'}$ reduces 
$M_{\rm SUSY}$ from about $1900~{\rm GeV}$ to about 
$1200~{\rm GeV}$. Note, in passing, that $\lambda_L$, 
in this case, is about 0.32.

\begin{figure}[t]
\centerline{\epsfig{file=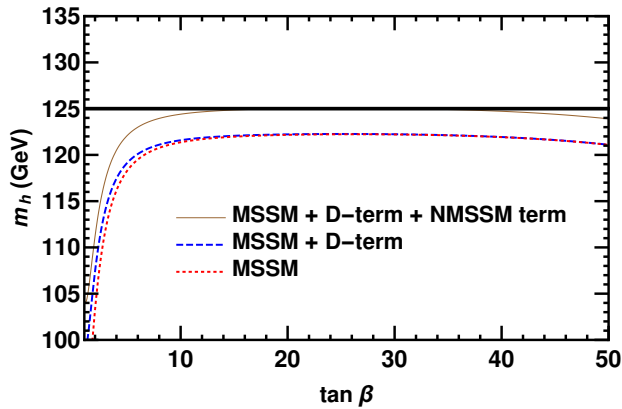,width=8.4cm}}
\caption{Higgs boson mass $m_h$ in the decoupling limit 
and for maximal stop quark mixing versus $\tan\beta$
for $M=10~{\rm TeV}$, $\tilde{\lambda}_{\mu}=0.3$, 
$M_{\rm SUSY}=1200~{\rm GeV}$, and $m_{3/2}=4~{\rm TeV}$.
The notation is the same as in Fig.~\ref{higgsmsusy}.}
\label{higgstanb}
\end{figure}

In Fig.~\ref{higgstanb}, we plot $m_h$ in the decoupling 
limit and for maximal stop quark mixing versus $\tan\beta$ 
for $M=10~{\rm TeV}$, $\tilde{\lambda}_{\mu}=0.3$, 
$M_{\rm SUSY}=1200~{\rm GeV}$, and $m_{3/2}=4~{\rm TeV}$.
We see that the experimental value of $m_h$ is achieved 
at $\tan\beta=20$ as it should consistently with 
Fig.~\ref{higgsmsusy}. However, as one can see from  
Fig.~\ref{higgstanb}, the observed $m_h$ can be 
practically obtained 
in a wide range of $\tan\beta$'s. Note that, without the 
inclusion of the D-term contribution from $U(1)_{\psi'}$, 
the Higgs boson mass remains well below its observed 
value for all the values of $\tan\beta$. This again shows 
the crucial role of the D-term for obtaining the observed 
value of $m_h$ with relatively low stop quark masses. 
Finally, we notice that, for larger $\tan\beta$'s, $m_h$ 
decreases as $\tan\beta$ increases in all three cases 
depicted in this figure. This is due to the relatively 
large value of $\mu$.      

\begin{figure}[t]
\centerline{\epsfig{file=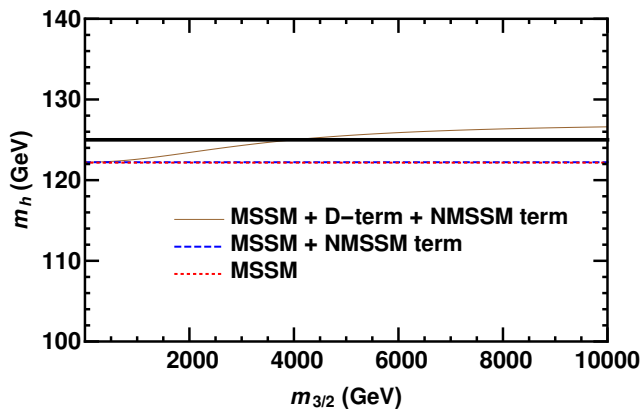,width=8.4cm}}
\caption{Higgs boson mass $m_h$ in the decoupling 
limit and for maximal stop quark mixing versus $m_{3/2}$
for $M=10~{\rm TeV}$, $\tilde{\lambda}_{\mu}=0.3$, 
$\tan\beta=20$, and $M_{\rm SUSY}=1200~{\rm GeV}$. The 
notation is the same as in Fig.~\ref{higgsmsusy}.}
\label{higgsm32}
\end{figure}

In Fig.~\ref{higgsm32}, we depict $m_h$ under the same 
assumptions versus $m_{3/2}$ for $M=10~{\rm TeV}$, 
$\tilde{\lambda}_{\mu}=0.3$, $\tan\beta=20$, and 
$M_{\rm SUSY}=1200~{\rm GeV}$. The observed Higgs 
boson mass is obtained at $m_{3/2}=4~{\rm TeV}$ 
consistently with Figs.~\ref{higgsmsusy} and 
\ref{higgstanb}. We see again that, without the 
D-term, $m_h$ remains well below its observed value
for all $m_{3/2}$'s. We also observe that, without the 
D-term, $m_h$ is independent from the value of 
$m_{3/2}$ as it should. 

In the present numerical example, the cross 
section of the diphoton excess in Eq.~(\ref{fb}) turns 
out to be equal to 1.94~{\rm fb}. Needless to say that 
higher cross sections can be obtained for higher values 
of $\kappa$. The diphoton re\-sonance 
mass, as already discussed, is equal to $6~{\rm TeV}$ 
and the diquark masses about $3~{\rm TeV}$. In conclusion, 
we see that our model can predict diphoton and diquark 
resonances which hopefully can be observed in future 
expe\-riments.  

\section{Sterile Neutrinos}
\label{sec:sterile}

After the spontaneous breaking of the $U(1)_{\psi'}$ 
symmetry, the 
fermionic components of the three superfields $N_i$, which 
are SM singlets, acquire masses $m_{N_i}\simeq\lambda_N^i 
M^2/m_{\rm P}$ via the last superpotential coupling in 
Eq.~(\ref{W}). These masses can be $\lesssim 0.1~{\rm eV}$ 
for $M\sim 10~{\rm TeV}$ and these fermionic fields, which 
are stable on account of the $Z_2'$ symmetry in 
Table~\ref{tab:fields}, can act as sterile neutrinos. 

In the early universe, the sterile neutrinos are kept in 
equilibrium via reactions of the sort $N_i\bar{N}_i
\leftrightarrow$ a pair of SM particles or $N_i+ 
{\rm a~SM~particle}\leftrightarrow N_i+{\rm a~SM~particle}$. 
These reactions proceed via a s- or t-channel exchange of 
a $Z'$ gauge boson. The thermal average $\langle\sigma v
\rangle$, where $\sigma$ is the corresponding cross section
and $v$ the relative velocity of the annihilating particles,
is estimated to be of order $T^2/M^4$ with $T$ being 
the cosmic temperature. The interaction rate per sterile 
neutrino is then given by 
\beq
\Gamma_{N_i}=n\vev{\sigma v}\sim \frac{T^5}{M^4},
\eeq
where $n\sim T^3$ is the number density of massless 
particles in thermal equilibrium. The decoupling 
temperature $T_{\rm D}$ of sterile neutrinos is estimated 
from the condition
\beq
\Gamma_{N_i}\sim H\sim \frac{T^2}{m_{\rm P}},
\eeq 
where $H$ is the Hubble parameter. This condition implies 
that   
\beq
T_{\rm D}\sim M \left(\frac{M}{m_{\rm P}}
\right)^{\frac{1}{3}}.
\label{TD}
\eeq
Here we followed the same strategy as the one used for 
estimating the SM neutrino decoupling temperature 
via processes involving weak gauge boson exchange. 
In the case of ordinary neutrinos, however, the scale 
$M$ should be identified with the electroweak scale, which 
is of order $100~{\rm GeV}$, and the decoupling temperature 
turns out to be of order $1~{\rm MeV}$. From Eq.~(\ref{TD}), 
we see that $T_{\rm D}$ scales like $M^{4/3}$. So, in our 
case and for $M\simeq 10~{\rm TeV}$, $T_{\rm D}$ is 
expected to be of order $460~{\rm MeV}$, which is well 
above the critical temperature for the QCD transition. 

The effective number of massless degrees of freedom in 
equilibrium right after the decoupling of sterile neutrinos 
is 61.75. At $T\sim 1~{\rm MeV}$ and just before the 
decoupling of the SM neutrinos, this number is reduced to 
10.75. So, due to entropy conservation in each comoving 
volume, the temperature of ordinary neutrinos $T_\nu$ is 
raised relative to the temperature of the sterile neutrinos 
$T_N$ by a factor $(61.75/10.75)^{1/3}$. Consequently, the
contribution of the three sterile neutrinos to the effective 
number of neutrinos at big bang nucleosynthesis is 
\beq
\Delta N_\nu=3\times \left(\frac{10.75}{61.75}
\right)^{\frac{4}{3}}\simeq 0.29.
\label{DNnu} 
\eeq
This result is perfectly compatible with the Planck satellite 
bound \cite{Neff} on the effective number of massless 
neutrinos
\beq
N_\nu=3.15\pm 0.23.
\eeq

Note that although the derivation of our estimate in 
Eq.~(\ref{DNnu}) is somewhat rough, we believe that the
result is quite accurate. This is due to the fact that 
the effective number of massless degrees of freedom in 
equili\-brium right after the decoupling of sterile 
neutrinos does not change if $T_{\rm D}$ varies between 
the critical temperature of the QCD transition, which is 
about $200~{\rm MeV}$, and the mass of the charm quark 
$m_c\simeq 1270~{\rm MeV}$. Also, a more accurate 
determination of the decoupling temperature of ordinary 
neutrinos does not change the effective number of 
massless degrees of freedom in equilibrium just before 
this temperature is reached. 

\section{Dark Matter}
\label{sec:dm}

The scalar component of the superfield $N_i$, which is 
expected to have mass of order $m_{3/2}$, can decay 
into a fermionic $N_i$ and a particle-sparticle pair 
via a $Z'$ gau\-gino exchange provided that this is
kinematically allowed. A necessary (but not sufficient) 
condition for this decay to be possible is that there 
exist sparticles which are lighter than the scalar 
$N_i$. Note that, as a consequence of the unbroken 
discrete symmetry $Z_2'$, the decay pro\-ducts of the 
scalar $N_i$ should necessarily contain an odd number 
of $N_j$ superfields.

If the decay of the lightest scalar $N_i$ (denoted as 
$\hat{N}$) is kinematically blocked, this particle can 
contribute to the cold dark matter in the universe. 
In the early universe, the scalar $\hat{N}$ is kept 
in equilibrium since, for example, a pair of these 
scalars can annihilate into a pair of SM particles 
via a $Z'$ gauge boson exchange. The thermal average 
$\langle\sigma v\rangle$ in this case and for s-wave 
annihilation is expected to be 
\beq 
\langle\sigma v\rangle\sim \frac{m_{\hat{N}}^2}{M^4},
\eeq 
where $m_{\hat{N}}$ is the mass of the scalar 
$\hat{N}$. 

Following the standard analysis of Ref.~\cite{kolb}, 
we can estimate the freeze-out temperature $T_{\rm f}$ 
of the sterile sneutrino $\hat{N}$ as well as its 
relic abundance $\Omega_{\hat{N}}h^2$ in the universe. 
To this end, we take $M\simeq 5.34~{\rm TeV}$, which 
saturates the lower bound on $m_{Z'}$ \cite{Zprime} 
mentioned in Sec.~\ref{sec:diphoton}. The 
requirement that $\Omega_{\hat{N}}h^2$ equals the cold 
dark matter abundance $\Omega_{\rm CDM}h^2\simeq 0.12$ 
from the Planck satellite data \cite{planck} then 
implies that $m_{\hat{N}}\simeq 1.25~{\rm TeV}$. The 
freeze-out temperature $T_{\rm f}$ in this case is about
$51~{\rm GeV}$ and the corresponding number of massless
degrees of freedom $86.25$. Higher values of $M$ require
even higher values of $m_{\hat{N}}$. So we see that the 
SUSY spectrum is pushed up considerably if the decay of 
the lightest sterile sneutrino is kinematically blocked 
and this particle contributes to the cold dark matter of 
the universe. 

The model possesses an accidental lepton parity symmetry
$Z_2^{\rm lp}$ under which the superfields $l$, $e^c$, 
$\nu^c$, $L$, $\bar{L}$ are odd. Combining this symmetry 
with the baryon parity $Z_2^{\rm bp}$ subgroup of 
$U(1)_B$ under which $q$, $u^c$, $d^c$ are odd, we 
obtain a matter parity symmetry $Z_2^{\rm mp}$ under 
which  $q$, $u^c$, $d^c$, $l$, $e^c$, $\nu^c$, $L$, 
$\bar{L}$ are odd. A discrete R-parity can then be 
generated if we combine this symmetry with fermion 
parity. The bosonic $q$, $u^c$, $d^c$, $l$, $e^c$, 
$\nu^c$, $L$, $\bar{L}$ and the fermionic $H_u^i$, 
$H_d^i$, $D_i$, $D^c_i$, $N_i$, $S$, $N$, $\bar{N}$ are
odd under this R-parity. Note that the decay products 
of these particles with the exception, of course, of 
the fermionic $N_i$ cannot contain a single $N_i$ 
because of the $Z_2'$ symmetry. Also, they cannot 
contain a single $L$, $\bar{L}$, $H_u^\alpha$, 
$H_d^\alpha$ except, of course, for the decay 
products of the bosonic $L$, $\bar{L}$ and fermionic 
$H_u^\alpha$, $H_d^\alpha$ themselves as a consequence 
of the $Z_2$ symmetry. The $S$, $N$, $\bar{N}$ fermions 
can decay into a Higgs boson-Higgsino pair, while 
the $D_i$, $D^c_i$ fermions can decay into a 
quark-squark pair. So all the particles with negative 
R-parity, except the bosonic $L$, $\bar{L}$ and the 
fermionic $H_u^\alpha$, $H_d^\alpha$, $N_i$ end up 
yielding the usual stable lightest sparticle of MSSM 
which can, in principle, participate in the cold dark 
matter of the universe.

The possible fate of the $N_i$ superfields has been
already discussed. The $Z_2$ symmetry and R-parity 
imply that the lightest state in the bosonic $L$,  
$\bar{L}$ and fermionic $H_u^\alpha$, $H_d^\alpha$, 
or in the fermionic $L$, $\bar{L}$ and bosonic 
$H_u^\alpha$, $H_d^\alpha$, which is hopefully 
neutral, is stable. We thus have two more candidates 
for cold dark matter. Their relic abundances in the 
universe depend on details. However, if their masses
are large, these abundances can be negligible. Finally,
let us mention that, if the breaking scale $\vev{N}$ of 
$U(1)_{\psi'}$ is increased to about $10^3~{\rm TeV}$, 
the sterile neutrinos become plausible candidates 
for keV scale warm dark matter (for a recent review see 
Ref.~\cite{keVsterile}). In conclusion, we see
that the model possesses many possible candidates for
the composition of dark matter. 
   
\section{Summary}
\label{sec:sum}

We have explored the implications of appending a $U(1)$ gauge symmetry 
to the MSSM gauge group $SU(3)_{\rm c}\times SU(2)_{\rm L}\times U(1)_Y$. 
This $U(1)$ symmetry, referred to here as $U(1)_{\psi'}$, arises from a 
linear combination of $U(1)_{\chi}$ and $U(1)_{\psi}$ contained in $E_6$. 
The three matter 27-plets in $E_6$ give rise to three $SO(10)$ singlet 
fermions $N_i$, called ste\-rile neutrinos, which are prevented from 
acquiring masses via renormalizable couplings by a combination of 
symmetries, especially a $U(1)$ R symmetry. Thus, for a re\-latively low 
($\sim 10~{\rm TeV}$ or so) breaking scale of $U(1)_{\psi'}$, these 
fermionic $N_i$'s, the lightest of  which happens to be stable, only 
acquire tiny 
masses $\lesssim 0.1~{\rm eV}$ and their contribution as fractional 
cosmic neutrinos during nucleosynthesis has been estimated. The lightest 
sterile sneutrino as well as two more particles, which are stable on 
account of discrete symmetries, can, under certain 
circumstances, be cold dark matter candidates in addition to the usual
lightest sparticle of MSSM. Note that 
the breaking of $U(1)_{\rm \psi'}$ at suitably higher energies, of 
order $10^{3}~{\rm TeV}$ or so, would yield keV scale masses for the 
fermionic $N_i$'s and thus transform them into plausible warm dark 
matter candidates. The D-term for $U(1)_{\psi'}$ can contribute 
appreciably to the mass of the lightest neutral CP-even MSSM Higgs 
boson. Consequently, the observed value of this mass 
can be obtained in the decoupling limit with relatively light stop 
quarks. The spontaneous breaking of $U(1)_{\rm \psi'}$ yields 
superconducting cosmic strings which presumably were not inflated 
away. The model also predicts the existence of diquark and diphoton 
resonances 
which may be found at the LHC or its future upgrades. The MSSM $\mu$
pro\-blem is naturally resolved. The right handed neutrinos can 
acquire large masses, which allows the standard seesaw mechanism
and the leptogenesis scenario to be realized. Baryon number is
conserved to all orders in perturbation theory rendering a stable 
proton.  

\acknowledgments{Q.S. thanks Nobuchika Okada for a clear explanation 
of the fractional cosmic neutrino contribution by the ste\-rile 
neutrinos. Q.S.and A.H. are supported in part by the DOE Grant 
DE-SC0013880.}

\def\ijmp#1#2#3{{Int. Jour. Mod. Phys.}
{\bf #1},~#3~(#2)}
\def\plb#1#2#3{{Phys. Lett. B }{\bf #1},~#3~(#2)}
\def\zpc#1#2#3{{Z. Phys. C }{\bf #1},~#3~(#2)}
\def\prl#1#2#3{{Phys. Rev. Lett.}
{\bf #1},~#3~(#2)}
\def\rmp#1#2#3{{Rev. Mod. Phys.}
{\bf #1},~#3~(#2)}
\def\prep#1#2#3{{Phys. Rep. }{\bf #1},~#3~(#2)}
\def\prd#1#2#3{{Phys. Rev. D }{\bf #1},~#3~(#2)}
\def\npb#1#2#3{{Nucl. Phys. }{\bf B#1},~#3~(#2)}
\def\np#1#2#3{{Nucl. Phys. B }{\bf #1},~#3~(#2)}
\def\npps#1#2#3{{Nucl. Phys. B (Proc. Sup.)}
{\bf #1},~#3~(#2)}
\def\mpl#1#2#3{{Mod. Phys. Lett.}
{\bf #1},~#3~(#2)}
\def\arnps#1#2#3{{Annu. Rev. Nucl. Part. Sci.}
{\bf #1},~#3~(#2)}
\def\sjnp#1#2#3{{Sov. J. Nucl. Phys.}
{\bf #1},~#3~(#2)}
\def\jetp#1#2#3{{JETP Lett. }{\bf #1},~#3~(#2)}
\def\app#1#2#3{{Acta Phys. Polon.}
{\bf #1},~#3~(#2)}
\def\rnc#1#2#3{{Riv. Nuovo Cim.}
{\bf #1},~#3~(#2)}
\def\ap#1#2#3{{Ann. Phys. }{\bf #1},~#3~(#2)}
\def\ptp#1#2#3{{Prog. Theor. Phys.}
{\bf #1},~#3~(#2)}
\def\apjl#1#2#3{{Astrophys. J. Lett.}
{\bf #1},~#3~(#2)}
\def\apjs#1#2#3{{Astrophys. J. Suppl.}
{\bf #1},~#3~(#2)}
\def\n#1#2#3{{Nature }{\bf #1},~#3~(#2)}
\def\apj#1#2#3{{Astrophys. J.}
{\bf #1},~#3~(#2)}
\def\anj#1#2#3{{Astron. J. }{\bf #1},~#3~(#2)}
\def\mnras#1#2#3{{MNRAS }{\bf #1},~#3~(#2)}
\def\grg#1#2#3{{Gen. Rel. Grav.}
{\bf #1},~#3~(#2)}
\def\s#1#2#3{{Science }{\bf #1},~#3~(#2)}
\def\baas#1#2#3{{Bull. Am. Astron. Soc.}
{\bf #1},~#3~(#2)}
\def\ibid#1#2#3{{\it ibid. }{\bf #1},~#3~(#2)}
\def\cpc#1#2#3{{Comput. Phys. Commun.}
{\bf #1},~#3~(#2)}
\def\astp#1#2#3{{Astropart. Phys.}
{\bf #1},~#3~(#2)}
\def\epjc#1#2#3{{Eur. Phys. J. C}
{\bf #1},~#3~(#2)}
\def\nima#1#2#3{{Nucl. Instrum. Meth. A}
{\bf #1},~#3~(#2)}
\def\jhep#1#2#3{{J. High Energy Phys.}
{\bf #1},~#3~(#2)}
\def\jcap#1#2#3{{J. Cosmol. Astropart. Phys.}
{\bf #1},~#3~(#2)}
\def\lnp#1#2#3{{Lect. Notes Phys.}
{\bf #1},~#3~(#2)}
\def\jpcs#1#2#3{{J. Phys. Conf. Ser.}
{\bf #1},~#3~(#2)}
\def\aap#1#2#3{{Astron. Astrophys.}
{\bf #1},~#3~(#2)}
\def\mpla#1#2#3{{Mod. Phys. Lett. A}
{\bf #1},~#3~(#2)}

\end{document}